\documentclass[aps,prl,reprint,groupedaddress]{revtex4-2}
\usepackage[top=2.5cm,bottom=2.5cm,right=3cm,left=3cm]{geometry}
\usepackage[utf8]{inputenc}
\usepackage{braket}
\usepackage{amsmath}
\usepackage{bbm}
\usepackage{appendix}
\usepackage{amsfonts}
\usepackage{comment}
\usepackage{cancel}
\usepackage{bbold}
\usepackage{graphicx}
\usepackage{float}
\usepackage{MnSymbol}
\usepackage{mathtools}
\usepackage{soul}
\usepackage{tabularx}
\usepackage{array}
\usepackage{accents}
\usepackage{bm}
\usepackage{placeins}
\usepackage{quantikz}

\newcolumntype{Y}{>{\centering\arraybackslash}X}

\usepackage[colorlinks, linkcolor=blue]{hyperref}
\usepackage[nameinlink, capitalise]{cleveref}

\newcommand{\mr}{\mathrm}

\newcommand{\mc}{\mathcal}

\newcommand{\bs}{\boldsymbol}

\newcommand{\NQL}{\ensuremath{{N_\mathrm{QL}}}}
\newcommand{\Ng}{\ensuremath{{N_\mathrm{g}}}}
\newcommand{\bigbox}{\mathop{\raisebox{-0.5ex}{\scalebox{1.5}{$\square$}}}\displaylimits}
\newcounter{definition}
\newtheorem{Definition}[definition]{Definition}

\newcounter{theorem}
\newtheorem{Theorem}[theorem]{Theorem}

\newcounter{proposition}
\newtheorem{Proposition}[proposition]{Proposition}

\begin{document}

	\title{Quantum-like product states constructed from classical networks}
	\author{Gregory D. Scholes}
	\email{gscholes@princeton.edu}
	\author{Graziano Amati}
	\affiliation{Department of Chemistry, Princeton University, Princeton New Jersey 08544 U.S.A.\looseness=-1}

	\date{\today}

	\begin{abstract}
	Can complex classical systems be designed to exhibit superpositions of tensor products of basis states, thereby mimicking quantum states? We exhibit a one-to-one map between the product basis of quantum states comprising an arbitrary number of qubits and the eigenstates of a construction comprising classical oscillator networks. Specifically, we prove the existence of this map based on Cartesian products of graphs, where the graphs depict the layout of oscillator networks. We show how quantum-like gates can act on the classical networks to allow quantum-like operations in the state space.
	\end{abstract}

	\maketitle
    \newpage

Quantum states encode information differently from classical states, allowing certain processes to be carried out more effectively, that is, using less resources, than is possible by using classical states---a fact known as the \textit{quantum advantage}. Underpinning the power of quantum states is their construction as superpositions of tensor products of the basis states\cite{Peres, Werner1989, Peres1996, Horodecki2009, NielsenChuang, Stormer2008}. However, it is a challenge to produce robust quantum states, especially incorporating many qubits, so as to leverage the quantum advantage. It is, moreover, difficult to imagine how a quantum advantage might be exploited in complex systems like large chemical or biological systems. 

The question motivating the present work is: Can complex classical systems be designed to exhibit superpositions of tensor products of basis states that mimic corresponding quantum states\cite{Khrennikov1, Khrennikov2018SciRep, Spreeuw1998}? The essence of the paper is proof of Proposition 1. 
The resulting insight enables a graph construction---which could be abstract or could be concretely associated to a classical network---to be mapped to a state space that mimics the state space enjoyed by quantum systems. The demonstration of this map is fascinating, and it might give insight into the interplay between classical and quantum systems. The result opens up  a way to take maps on the state space (e.g. quantum gates) and map them back to maps on the graph, which represents a physical, classical system. 

In recent work, we explored how special networks of oscillators, based on expander graphs, can be constructed so as to generate a discrete emergent state that represents a superposition of two states of coupled oscillator sub-networks\cite{ScholesQLstates}. We called that the `quantum-like' (QL) bit. The key advances of our prior studies are: (i) to demonstrate how the properties of the expander graph network enable the system to be remarkably resistant to various kinds of disorder that would otherwise contribute to decoherence (see ref. \cite{ScholesEntropy}), and (ii) to exhibit a two-state system that serves as a QL bit. 

In the present work we make a significant extension to this work by showing explicitly how to scale the QL states to accommodate an arbitrary number of QL bits. The states associated with this construction have product states as their basis, thereby building correlations into the basis that enable the superposition states to have special properties, similar to those of quantum systems. Specifically, the quantum mechanical state space needed to encode all the information associated to a system with many qubits is built from the tensor product of the Hilbert spaces of each qubit in a system, $\mathcal{H} = \mathcal{H}_{1} \otimes \mathcal{H}_{2} \otimes \cdots \otimes \mathcal{H}_{n}$. One possible basis for describing the state of a quantum system can then be produced by setting the qubits in either the $|0\rangle$ or $|1\rangle$ state, to obtain $2^{n}$ unique product states in $\mathcal{H}$. This generates a basis of tensor products. The general states can comprise superpositions of product states---states of special interest because to date they have no classical counterpart when endowed with the additional property of nonlocality.  

Here we provide an explicit construction of superpositions of product states of QL bits. We do this by producing new graphs formed from Cartesian products of an arbitrary number of QL bits. The emergent eigenstates of these product graphs are  tensor products of the emergent states of the constituent QL bits. 

A QL-bit is constructed from a pair of \textit{interacting} networks, or subgraphs. The model is abstract, but one way to make it concrete is to define the nodes of the network (vertices in the graph representation) be be classical oscillators and the links between nodes (edges in the graph) are couplings between the oscillators. We thus represent the map of how the oscillators in a network are coupled to each other using graphs, and the properties of those graphs tell us about the properties of the corresponding network. A graph $G(n,m)$, that we often write simply as $G$, comprises $n$ vertices and a set of $m$ edges that connect pairs of vertices. The size of a graph or subgraph, that is, the number of vertices, is written $|G|$. The spectrum of a graph $G$ is defined as the spectrum (i.e. eigenvalues in the case of a finite graph) of its adjacency matrix $A$. For background see \cite{Diestel,Janson2000, Bollobas2001}. 

For the QL bits, we suggested that each of the subgraphs is an expander graph, that is, they have a special form of connectivity among the vertices  that guarantees the highest eigenvalue of each graph is associated with an emergent state---a state isolated from all other eigenstates of the graph\cite{ScholesQLstates}.  We refer to the emergent state as the state of largest eigenvalue, consistent with standard work on spectra of graphs. However, in practice, by setting the sign of coupling-edge entries in the adjacency matrix to negative values, then the emergent state is the least eigenvalue\cite{ScholesQLstates}.

The spectral isolation enabled by the emergent state protects it from decoherence that could be caused by energetic or structural disorder, as shown in refs. \cite{ScholesEntropy, Scholes2020}. Families of graphs with this property are known as expander graphs\cite{Sarnak2004, expandersguide, Lubotzky, Expanders, Expanders2, Alon1986}. The $d$-regular graphs include prototypical expander graphs; here every vertex is adjacent to $d$ other vertices in the graph:
\begin{Definition}
	($d$-regular graph) A graph $G$ is $d$-regular if every vertex has degree (valency) $d$. That is, every vertex connects to $d$ edges.
\end{Definition}

Spectra of $d$-regular graphs on $n$ vertices\cite{McKay1981} comprise two distinct features: the emergent state with eigenvalue $d$ and a set of `random states' with eigenvalues in the interval $[-2\sqrt{d-1}, 2\sqrt{d-1}]$.  For $d$-regular graphs with large $d$, the optimal gap between the first and second eigenvalues ($\lambda_0$ and $\lambda_1$ respectively) can be substantial, and is set by the Alon-Boppana bound\cite{Alon1986, Nilli1991, Friedman2008, HMY2024, expandersguide}: 
\begin{Theorem} 
	(Alon-Boppana) Let $G$ be a $d$-regular graph, then the second largest eigenvalue satisfies
	\begin{equation}
		\lambda_1 \ge 2\sqrt{d - 1} - o_n(1).
	\end{equation}
\end{Theorem}

To produce the QL-bit, define two $d$-regular graphs that are denoted basis graphs, labelled, for example, $G_{a1}$ and $G_{a2}$. These basis graphs are coupled to each other by randomly adding a small fraction of edges from vertices in $G_{a1}$ to $G_{a2}$. In the present paper we do this by randomly adding edges with probability $p$, set to 0.1 or 0.2. The resulting emergent states are the in- and out-of-phase linear combinations of the emergent states of the base graphs. We write the emergent eigenstate of $G_{a1}$ as $a_1$ and that of $G_{a2}$ as $a_2$. Hence we write the emergent states of the QL bit as the states $a_1 + a_2$. and $a_1 - a_2$. In this notation, $a_1 \pm a_2$ is not an algebraic sum. It denotes the coefficients of the eigenstates that are divided into two sets, one associated with subgraph $G_{a1}$, $a_1$, and one with subgraph $G_{a2}$, $a_2$.

In prior work\cite{ScholesQLstates} we showed that QL bits can be connected pairwise to produce arbitrary new states, QL states. Here we propose a general construction using products of QL bit graphs. We exhibit a one-to-one map between the tensor product basis of states of two-level systems (the basis described above used for systems of qubits) and the states generated by Cartesian products\cite{GraphProducts} of QL bits. 

\begin{Definition}
	(Cartesian product of graphs) $G \Box H$ is defined on the Cartesian product of vertex sets, $V(G) \times V(H)$. Let $\{u, v, \dots\} \in V(G)$ and $\{x, y \dots\} \in V(H)$. Let $E(G)$ and $E(H)$ be the set of edges in $G$ and $H$ respectively. The edge set of the product graph $G \Box H$ is defined with respect to all edges in $G$ and all edges in $H$  as follows. We have an edge in $G \Box H$ when
	\begin{align*}
		\text{either} \quad [u,v] \in E(G) \; \textrm{and} \; x = y \\
		\text{or} \quad [x,y] \in E(H) \; \textrm{and} \; u = v.
	\end{align*}
\end{Definition}

Some properties of the product might be useful. The Cartesian product of graphs $G \Box H$ is connected if and only if both $G$ and $H$ are connected. The Cartesian product of graphs is associative, i.e., for three graphs $G$, $H$ and $K$, $(G\Box H)\Box K \cong G\Box (H \Box K)$, where $\cong$ denotes an equivalence after index relabeling (i.e. an isomorphism).

Thus, we have an explicit mapping from the usual product basis of the Bell states to the basis graphs $G_{a1}$ and $G_{a2}$ of QL bit A and $G_{b1}$ and $G_{b2}$ of QL bit B as follows:
\begin{Proposition}
    \begin{align*}
	     G_{a1} \Box G_{b1} \quad \rightarrow \quad |0\rangle_A |0\rangle_B  \\
	     G_{a1} \Box G_{b2} \quad \rightarrow \quad |0\rangle_A |1\rangle_B  \\
	     G_{a2} \Box G_{b1} \quad \rightarrow \quad |1\rangle_A |0\rangle_B  \\
	     G_{a2} \Box G_{b2} \quad \rightarrow \quad |1\rangle_A |1\rangle_B 
    \end{align*} 
    The map indicated by the arrow means that the graph eigenvector corresponding to the emergent state represents the product basis state.  Repeated application of the Cartesian product generates maps for any number of QL bits.
\end{Proposition}

The map is made explicit by a projection onto a new basis that we now describe. The eigenstates of the graphs are determined with respect to the basis of the (many) vertices of the graph. In order to describe the usual quantum states space for interacting two-level systems, we define a map from the eigenvectors of each QL bit to  an orthonormal basis of two-state systems in terms of the Hilbert space of each QL bit. The graph product states can thereby be related to the product states of a system of two-level entities. We have an orthonormal basis for the QL state $\{ u_1, u_2, \dots, u_n, x_1, x_2, \dots, x_k \}$, which defines a complete basis in $\mathbbm{R}^{(n+k)}$. Subgraph $G_{a1}$ with $n$ vertices is associated to basis functions labeled $u_1, u_2, \dots, u_n$ and subgraph $G_{a2}$ with $k$ vertices is associated to basis functions labeled $x_1, x_2, \dots, x_k$. For the emergent eigenstate of the graph we define the vectors $U_{a1} = c_1u_1 + c_2u_2 + \dots + c_nu_n$ and $X_{a2} = d_1x_1 + d_2x_2 + \dots + d_nx_k$, where $c_i$ and $d_i \in \mathbb{R}$ are coefficients from the graph eigenvector. By construction, $U_{a1}$ and $X_{a2}$ are orthogonal.

We define two orthogonal vectors, $J_{a1} = \frac{1}{\sqrt{n}}(u_1 + u_2 + \dots + u_n)$ and  $J_{a2} = \frac{1}{\sqrt{k}}(x_1 + x_2 + \dots + x_k)$. $J_{a1}$ is the basis vector corresponding to the state $|0\rangle_A$ of the two-level representation of QL bit A, and $J_{a2}$ is the basis vector corresponding to the state $|1\rangle_A$.  Hence we can project $G_A\Box G_B \rightarrow (U_{a1} + X_{a2}) \otimes (U_{b1} + X_{b2})$ to $\alpha_{00} |0\rangle_A \otimes |0\rangle_B + \alpha_{01} |0\rangle_A \otimes |1\rangle_B + \dots$, where the $\alpha_{ij}$ are coefficients, by evaluating $\alpha_{00} = \langle U_{a1}, J_{a1} \rangle \langle U_{b1}, J_{b1} \rangle $, $\alpha_{01} = \langle U_{a1}, J_{a1} \rangle \langle X_{b2}, J_{b2} \rangle $, etc. Here $\langle \; , \;  \rangle$ means inner product. Using this method we can project any eigenstate of arbitrary graph products to an explicit tensor product state on the basis of $G_{a1} \rightarrow |0\rangle_A$, $G_{a2} \rightarrow |1\rangle_A$, $G_{b1} \rightarrow |0\rangle_B$, etc.

Now to prove Proposition 1, we simply need to define the spectrum of the Cartesian product (see, for instance \cite{Barik2018}).

\begin{Proposition}\label{eq:eig_prod}
	(Spectrum of a Cartesian product of graphs) Given
	\begin{enumerate}
		\item[] A graph $G$, for which its adjacency matrix $A_G$ has eigenvalues $\lambda_i$ and eigenvectors $X_i$, and
		\item[] A graph $H$, for which its adjacency matrix $A_H$ has eigenvalues $\mu_i$ and eigenvectors $Y_i$, then
	\end{enumerate}
	the spectrum of $G \Box H$ contains eigenvalues $\lambda_i + \mu_j$ and the corresponding eigenvectors are $X_i \otimes Y_j$.
\end{Proposition}
The validity of Proposition 1 is also evident from the structure of the adjacency matrix of the Cartesian product $G\Box H$ between two graphs $G$ and $H$ of size $n$ and $k$, respectively, given by 
\begin{equation}
A_{G\Box H} = A_G\otimes I_n + I_k\otimes A_H,	
\end{equation}
where $I_m$ is the identity matrix of size $m$.

Arbitrary linear combinations of product states are produced when the basis graphs, for example $G_{a1}$ and $G_{a2}$, are coupled to produce the superposition states described above, $G_{a1 + a2}$, etc. For instance, we can produce the Bell states from graphs by suitable linear combinations.

\begin{figure*}
	\includegraphics[width=13.5 cm]{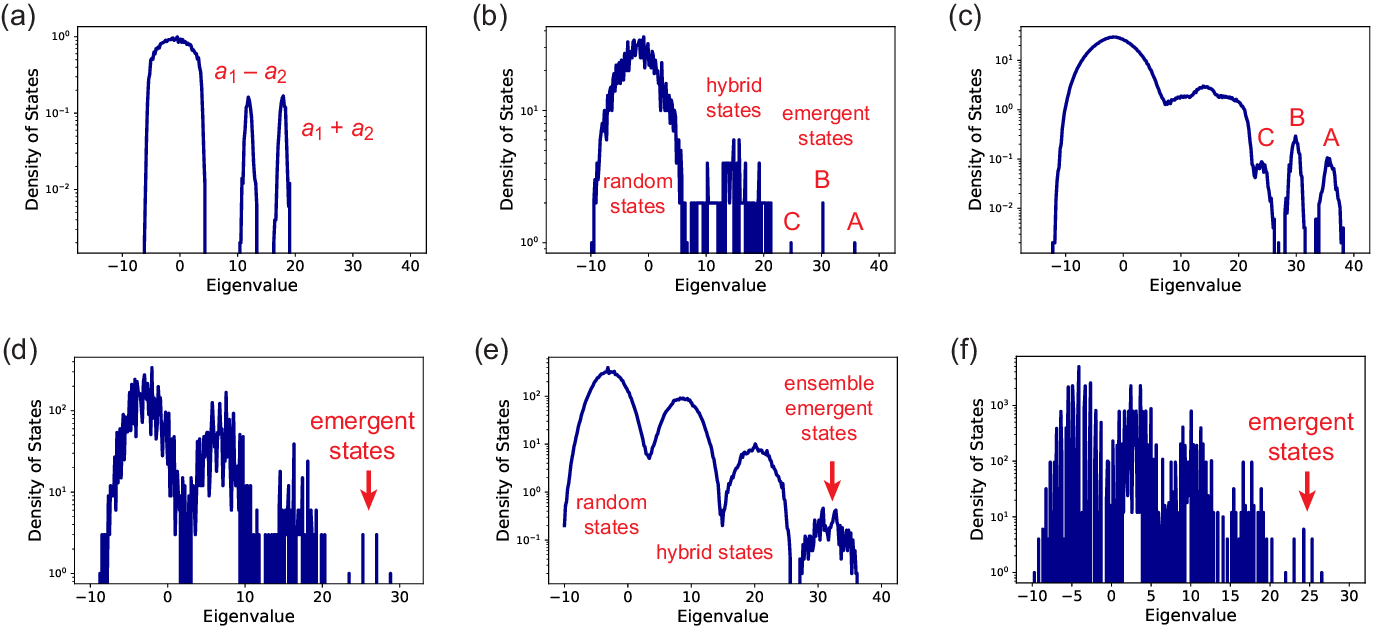}
	\caption{(a) The ensemble spectrum of one QL bit, composed of two $d$-regular basis graphs, showing the pair of emergent states (for each $d$-regular basis graph $n = 20$, $d = 15$, $m = 150$, $p = 0.2$). (b) The spectrum of a single Cartesian product graph of two identical QL bits, parameters as above. (c) The ensemble spectrum of Cartesian product graphs of two QL bits. The four emergent states are labeled A, B, C (see text). (d) The spectrum of a single Cartesian product graph of three identical QL bits (for each $d$-regular basis graph $n = 10$, $d = 9$, $m = 90$, $p = 0.1$). (e) The ensemble spectrum of Cartesian product graphs of three non-identical QL bits ($n = 12$, $d = 11$, $m = 132$, $p = 0.1$). The eight emergent states are indicated. (f) The spectrum of a single Cartesian product graph of four identical QL bits comprising 38,416 vertices. The 16 emergent states are indicated ($n = 7$, $d = 6$, $m = 45$, $p = 0.1$).   \label{fig4}}
\end{figure*}

In Figure~\ref{fig4} we show spectra of representative QL states formed from products of QL bit states. The two QL bits are each set up so that QL bit A has a highest emergent state $a_1 + a_2$ and for QL bit B we have $b_1 + b_2$. Each subgraph of the QL bits (e.g. $G_{a1}$) comprises 20 vertices and is a $d$-regular random graph with $d = 15$. Edges connect the subgraphs with probability $p$, noted in the figure caption. This random addition of connecting edges introduces disorder in the spectrum. That is evident in the ensemble calculation for the spectrum of QL bit A, Figure~\ref{fig4}a. This spectrum of one QL bit, composed of two coupled $d$-regular basis graphs, comprises two emergent states with eigenvalue $d$ split or `repelled' by the coupling of magnitude $\Delta = n_c/n$, where $n_c$ is the number of coupling edges connecting the subgraphs\cite{ScholesQLstates}.

In Figure~\ref{fig4}b we show an example of one spectrum for the QL state, that is, the spectrum of the graph product. The emergent state with highest eigenvalue, labelled A, is associated with $G_{a1 + a2} \Box G_{b1 + b2}$ that projects to $|0\rangle_A |0\rangle_B + |0\rangle_A |1\rangle_B + |1\rangle_A |0\rangle_B +  |1\rangle_A |1\rangle$.  The other states are also contained in this spectrum: B labels the degenerate states from $G_{a1 - a2} \Box G_{b1 + b2}$ and $G_{a1 + a2} \Box G_{b1 - b2}$, while C labels $G_{a1 - a2} \Box G_{b1 - b2}$. By changing the signs of the edges in the graphs, we can associate any of these states with the highest emergent eigenvalue. The corresponding ensemble spectrum is shown in Figure~\ref{fig4}c. These spectra display four emergent states. One state, A, has eigenvalue $d_a + \Delta_a  + d_b + \Delta_b$, two states, B, have eigenvalues $d_a - \Delta_a  + d_b + \Delta_b$ and $d_a + \Delta_a  + d_b - \Delta_b$, and one state, C, is located at $d_a - \Delta_a  + d_b - \Delta_b$. We extend the numerical results to display spectra of graph products of three and four QL-bits, Figure~\ref{fig4}d-f. The random states, hybrid states, and emergent states are various linear combinations of QL bit eigenvalues.

The spectrum of a Cartesian product of $N$ QL bits can include all $2^N$ emergent eigenvalues, although it is likely that the lower eigenvalues will be hidden by the random state spectrum. Assuming for simplicity that all the QL bit basis graphs are $d$-regular and that each splitting is $\Delta$ then the greatest eigenvalue is $Nd + N\Delta$. It is separated from the next highest eigenvalue by $2\Delta$, regardless of the value for $N$. The non-random part of the spectrum is centered at $Nd$ and has a spectral width of $[-N\Delta, +N\Delta]$. 

In terms of physical scaling of the system of QL bits, the number of oscillators increases exponentially with the number of QL bits in the product. This is because the product graph maps directly to the state space. For QL bits each comprising $n$ total vertices, the total number of vertices in a full product graph of $N$ QL bits is $n^N$. Owing to the design of the QL bits, it turns out that there is a more optimal product graph construction that gives a physical scaling of $n2^N$ and lifts to the full product graphs described here. .

To illustrate how QL states might provide a feasible resource for computation or function that is intermediate between the classical and quantum limits, we now show how to produce a quantum-like (QL) representation of highly entangled Greenberger–Horne–Zeilinger (GHZ) states \cite{erhard2020}, within a controlled approximation that requires only sub-exponential overhead in graph resources. This highlights the capability of the QL formalism to capture properties that extend beyond classical frameworks. 

To exploit QL states, we treat the graphs that compose them as a resource, $G=\bigbox_{q=1}^\NQL G^{(q)}$, defined as the Cartesian product of $\NQL$ QL-bits, where QL bit $q$ is represented by a graph $G^{(q)}$.
The adjacency matrix of $G$ is
\begin{equation}\label{eq:R}
	\mc R^{[\NQL]} = \sum_{q=1}^\NQL \mathbb{1}_{2\Ng}^{\otimes (q-1)}\otimes \mc R^{(q)}\otimes  \mathbb{1}_{2\Ng}^{\otimes (\NQL-q)},
\end{equation}
Each of the one-QL-bit graph resources is described by a $2\Ng$-dimensional adjacency matrix (i.e. each subgraph comprises $N_g$ vertices), with structure 
\begin{equation}\label{eq:Rq}
	\mc R^{(q)} = \begin{pmatrix} A^{(q)} & C^{(q)} \\ C^{(q)} & A^{(q)}\end{pmatrix}.
\end{equation}
$A^{(q)}$ describes a simple regular graph with valency $k$, while $C^{(q)}$ is a symmetric non-simple graph (with loops) and valency $l$.
The principal eigenvector of this resource, corresponding to the emergent state with the largest eigenvalue, is $\psi_{\bs 0}^{[\NQL]} = \bigotimes_{q=1}^\NQL\psi_{0}^{(q)}$. 

In Ref.~\cite{qlbits}, we proposed to build the QL analogy of an arbitrary quantum gate $V_{\mr g}$ [in SU(2)] by defining a transformation $U_{\mr{c.b.}}$ mapping the emergent states to the computational basis (c.b.).
	For example, for a single QL-bit $q$,
	\begin{equation}\label{eq:Ucb_1}
		U_{\mr{cb}}\psi_0^{(q)} = \begin{pmatrix}  1_\Ng \\  0_\Ng \end{pmatrix}, \hspace{5mm} U_{\mr{cb}}\psi_1^{(q)} = \begin{pmatrix}  0_\Ng \\  1_\Ng \end{pmatrix},
	\end{equation}
	where $x_M = (x,\cdots, x)/\sqrt{M} \in \mathbb C^{M}$, $x \in \mathbb C,\; M \in \mathbb N$.
	This led us to the definition of QL gates
	\begin{equation}\label{eq:Ug}
		U_{\mr g} =  U_{\mr{cb}}^{-1} (V_{\mr g} \otimes  \mathbb 1_{\Ng}) U_{\mr{cb}}.	
	\end{equation}

A notion of entanglement can be introduced in our approach by transforming the graph resources such that the principal eigenvectors are mapped as follows:
\begin{equation}\label{eq:Phi+_gen}
	\psi_{\bs 0}^{[\NQL]} \mapsto \frac 1 {\sqrt 2} \left(\psi_{\bs 0}^{[\NQL]} +\psi_{\bs 1}^{[\NQL]} \right)\equiv \Psi_{\mr{GHZ}}^{[\NQL]}.
\end{equation}
\Cref{eq:Phi+_gen} corresponds, in an abstract Hilbert space of $\NQL$ qubits, to the maximally-entangled GHZ state $\ket {\Psi_{\mr{GHZ}}} = \tfrac 1 {\sqrt 2} (\ket{0,\cdots,0} + \ket{1,\cdots,1})$.
$\Psi_{\mr{GHZ}}^{[\NQL]}$ is the principal eigenvector of the gate-transformed resource $\mc R^{[\NQL]}_{\mr{GHZ}} = U_{\mr{GHZ}}\mc R^{[\NQL]}U_{\mr{GHZ}}^\dagger$, where
\begin{equation}\label{eq:U_GHZ}
	U_{\mr{GHZ}} =U_{\mr {CNOT}}^{(1,\NQL)}\cdots  U_{\mr {CNOT}}^{(1,3)} U_{\mr {CNOT}}^{(1,2)} U_{\mr H}^{(1)}.
\end{equation}
The simple circuit associated to \cref{eq:U_GHZ} is illustrated in \cref{fig:U_GHZ}.
\begin{figure}[t]
	\centering
	\centering\begin{quantikz}
		\lstick{$\psi_{0}^{(1)}$} & \gate{U_{\mr H}^{(1)}} & \ctrl{1}  & \ctrl{2} & \ctrl{4} & \qw \\
		\lstick{$\psi_{0}^{(2)}$} & \qw      & \targ{}  & \qw      & \qw      & \qw \\
		\lstick{$\psi_{0}^{(3)}$} & \qw      & \qw      & \targ{}  & \qw      & \qw \\
		... &      &       &   ...    &  & ...\\
		\lstick{$\psi_{0}^{(\NQL)}$} & \qw      & \qw      & \qw      & \targ{}  & \qw
	\end{quantikz}
	\caption{QL representation of the transformation \cref{eq:Phi+_gen}, mapping the emergent state $\psi_{\bm 0}^{[\NQL]}$ onto the QL representation of the maximally-entangled GHZ state.}\label{fig:U_GHZ} 
\end{figure}
Here, $U_{\mr H}^{(1)} $ is defined by solving \cref{eq:Ucb_1} for $U_{\mr{c.b.}}$ (see Appendix A of Ref.~\cite{qlbits}), and by choosing $V_{\mr g} = V_{\mr H} = \tfrac 1 {\sqrt 2} \begin{pmatrix} 1 & 1 \\ 1 & -1 \end{pmatrix}$ in \cref{eq:Ug}.
	For the special case of regular resources \cref{eq:Rq}, this gate is simply
	$U_{\mr H}^{(1)} = V_{\mr H} \otimes \mathbb 1_{\Ng}$. The other terms in \cref{eq:U_GHZ} are
\begin{multline}\label{eq:UCNOT}
	U_{\mr {CNOT}}^{(q,p)} = \mathbb 1_{2\Ng}^{\otimes(p-1)}\otimes \mc P_0^{(p)}\otimes \mathbb 1_{2\Ng}^{\otimes(\NQL-p)} + \\ \mathbb 1_{2\Ng}^{\otimes(p-1)}\otimes \mc P_1^{(p)}\otimes \mathbb 1_{2\Ng}^{\otimes(q-p-1)}\otimes U_{\mr{X}}^{(q)}\otimes \mathbb 1_{2\Ng}^{\otimes(\NQL-q)},
\end{multline}
and correspond to the QL representation of the controlled-NOT (CNOT) gate.
In particular, the QL mapping of the Pauli-X (NOT) gate, $U_{\mr X}^{(q)} = \begin{pmatrix} 1 & 0 \\ 0 & -1\end{pmatrix}\otimes \mathbb 1_{\Ng}$, is applied to the target QL-bit $q$ depending on the state of the control term $p$.
The operators
\begin{equation}
	\mc P_{\sigma}^{(p)}= \frac 12 \left(\mathbb 1_{2\Ng} +(-1)^\sigma U_{\mr Z}^{(p)}\right), \hspace{10mm} \sigma \in \{0,1\},
\end{equation}
denote the projectors onto the emergent states of the control, and are expressed in terms of the QL map of the Pauli-Z gate, $U_{\mr Z}^{(p)}= \begin{pmatrix} 0 & 1 \\ 1 & 0\end{pmatrix}\otimes \mathbb 1_{\Ng}$. Note the the c.b. mapping \cref{eq:Ug} induces a ``symmetric'' definition of the Pauli -X and -Z gates in our approach.
In Ref.~\cite{qlbits}, we discussed that a significant reduction in computational complexity can be achieved within our approach if gate transformations preserve the structure of the Cartesian product. When this condition is met, individual resources can be allocated for the graph described in \cref{eq:R}, with a total storage cost of $2\Ng\NQL$. This allows processing to occur without the need to explicitly track the exponentially many edges in a fully connected graph of size $(2\Ng)^\NQL$, significantly reducing computational demands.

One-QL-bit gates, such as $U_{\mr H}^{(1)}$ in \cref{eq:U_GHZ}, preserve the symmetric structure of \cref{eq:R} exactly. On the other hand, CNOT gates \emph{approximately} preserve the Cartesian product structure, with a controlled error scaling as $\mc O(\sqrt{l/k})$ (as detailed in Appendix D of Ref.~\cite{qlbits}). 
In essence, the final resource upon information processing can be approximated as
\begin{multline}\label{eq:R_GHZ}
	\mc R^{[\NQL]}_{\mr{GHZ}} = \sum_{q=1}^\NQL \mathbb{1}_{2\Ng}^{\otimes (q-1)}\otimes \tilde{\mc R}_{\mr{GHZ}}^{(q)}\otimes   \mathbb{1}_{2\Ng}^{\otimes (\NQL-q)} \\ + \mc O\left(\NQL\sqrt{l/k}\right).
\end{multline}
where $\tilde{\mc R}_{\mr{GHZ}}^{(q)}$ are generally directed non-simple graphs.
Note that the error in \cref{eq:R_GHZ} can be minimized by choosing $l=l_{\mr{min}}=1$, $k=k_{\mr{max}} = \Ng-1$ and $\Ng\gg 1$.
Convergence to exact entangled states can be achieved through two distinct approaches: (i) by increasing the size of the basis graphs to infinity ($\Ng\to+\infty$), or (ii) by applying an exact transformation of the resources, which incurs exponential complexity in $\NQL$. Our approach strikes a balance between these extremes, offering a  ``middle ground'' that seeks to maximize accuracy in QL entanglement while minimizing computational costs on classical resources.

Superpositions of product states are one key reason for the special properties of quantum systems, including entanglement and nonlocality. The specific advance of the present work is that we have exhibited a one-to-one map between the product basis of quantum states, states in the state space comprising the tensor product of the Hilbert spaces of each qubit in a system, $\mathcal{H} = \mathcal{H}_{1} \otimes \mathcal{H}_{2} \otimes \cdots \otimes \mathcal{H}_{n}$, and the eigenstates of a construction comprising classical oscillator networks. We have proved the existence of this map based on Cartesian products of graphs, where the graphs depict the layout of the oscillator networks. We have thus shown that a state space comprising superpositions of product states---the hallmark of a quantum states---can be generated from classical systems.   The complex structure of these graphs (networks) gives a physical basis from which to understand the power of the tensor product basis of quantum states and thus may provide a framework to elucidate quantum correlations or yield insight into quantum phenomena.

\begin{acknowledgments}
This research was funded by the National Science Foundation under Grant No. 2211326 and the Gordon and Betty Moore Foundation through Grant GBMF7114.
\end{acknowledgments}


\vspace{6pt} 




\bibliography{product_states}

\end{document}